\documentclass[aps,prl,twocolumn,superscriptaddress,floatfix]{revtex4}
\usepackage{graphicx}
\usepackage{amssymb}

\begin{document}

\title{Resonant Spin Excitation in the High Temperature Superconductor Ba$_{0.6}$K$_{0.4}$Fe$_2$As$_2$}

\author{A. D. Christianson}
\affiliation{Neutron Scattering Science Division, Oak Ridge National Laboratory, Oak Ridge, TN 37831, USA}

\author{E. A. Goremychkin}
\affiliation{Materials Science Division, Argonne National Laboratory, Argonne, IL 60439-4845, USA}
\affiliation{ISIS Pulsed Neutron and Muon Facility, Rutherford Appleton Laboratory, Chilton, Didcot OX11 0QX, United Kingdom}

\author{R. Osborn}
\affiliation{Materials Science Division, Argonne National Laboratory, Argonne, IL 60439-4845, USA}
\email{ROsborn@anl.gov}

\author{S. Rosenkranz}
\affiliation{Materials Science Division, Argonne National Laboratory, Argonne, IL 60439-4845, USA}
\author{M. D. Lumsden}
\affiliation{Neutron Scattering Science Division, Oak Ridge National Laboratory, Oak Ridge, TN 37831, USA}
\author{C. D. Malliakas}
\affiliation{Materials Science Division, Argonne National Laboratory, Argonne, IL 60439-4845, USA}
\affiliation{Department of Chemistry, Northwestern University, Evanston, IL 60208-3113}
\author{l. S. Todorov}
\author{H. Claus}
\author{D. Y. Chung}
\affiliation{Materials Science Division, Argonne National Laboratory, Argonne, IL 60439-4845, USA}
\author{M. G. Kanatzidis}
\affiliation{Materials Science Division, Argonne National Laboratory, Argonne, IL 60439-4845, USA}
\affiliation{Department of Chemistry, Northwestern University, Evanston, IL 60208-3113}
\author{R. I. Bewley}
\author{T. Guidi}
\affiliation{ISIS Pulsed Neutron and Muon Facility, Rutherford Appleton Laboratory, Chilton, Didcot OX11 0QX, United Kingdom}

\begin{abstract}
The recent observations of superconductivity at temperatures up to 55K in compounds containing layers of iron arsenide \cite{Kamihara:2008p7994,Takahashi:2008p7382,Ren:2008p8174,Rotter:2008p8535} have revealed a new class of high temperature superconductors that show striking similarities to the more familiar cuprates.  In both series of compounds, the onset of superconductivity is associated with the suppression of magnetic order by doping holes and/or electrons into the band \cite{DeLaCruz:2008p8095} leading to theories in which magnetic fluctuations are either responsible for or strongly coupled to the superconducting order parameter \cite{Norman:2007p2671}.  In the cuprates, theories of magnetic pairing have been invoked to explain the observation of a resonant magnetic excitation that scales in energy with the superconducting energy gap and is suppressed above the superconducting transition temperature, T$_c$.  Such resonant excitations have been shown by inelastic neutron scattering to be a universal feature of the cuprate superconductors \cite{Hufner:2008p9494}, and have even been observed in heavy fermion superconductors with much lower transition temperatures \cite{Sato:2001p11225,Stock:2008p6655,Stockert:2008p11663}.  In this paper, we show neutron scattering evidence of a resonant excitation in Ba$_{0.6}$K$_{0.4}$Fe$_2$As$_2$, which is a superconductor below 38\,K \cite{Rotter:2008p8535}, at the momentum transfer associated with magnetic order in the undoped compound, BaFe$_2$As$_2$, and at an energy transfer that is consistent with scaling in other strongly correlated electron superconductors.  As in the cuprates, the peak disappears at T$_c$ providing the first experimental confirmation of a strong coupling of the magnetic fluctuation spectrum to the superconducting order parameter in the new iron arsenide superconductors. 
\end{abstract}

\date{\today}

\maketitle

Unconventional superconductivity has been the subject of considerable theoretical and experimental interest since the discovery of superconductivity in CeCu$_2$Si$_2$ and other heavy fermion compounds \cite{Steglich:1979p11524}, an interest that was only intensified by the discovery of cuprate superconductors with transition temperatures in excess of 100\,K \cite{Norman:2007p2671}.  Although significant progress has been made, the origin of unconventional superconductivity is still not understood.  The observation of a magnetic resonance in the spin excitation spectrum which appears concurrently with the onset of superconductivity in \textit{both} the high T$_c$ cuprates \cite{RossatMignod:1991p10509,Mook:1993p10706,Fong:1999p11512,Dai:2000p10619,He:2002p11430} and the heavy fermion superconductors \cite{Sato:2001p11225,Stock:2008p6655,Stockert:2008p11663} offers the tantalizing possibility of a unifying theme for unconventional superconductivity that spans a diverse range of superconducting materials.  Recently, a new family of superconductors containing layers of Fe$_2$As$_2$ has been discovered with T$_c$s in excess of 50\,K stimulating considerable experimental and theoretical activity \cite{Kamihara:2008p7994,Takahashi:2008p7382,Ren:2008p8174}. Although there is mounting evidence that the superconductivity in this new family is also unconventional \cite{Mazin:2008p8170}, there is as yet no consensus concerning the mechanism giving rise to superconductivity or even the superconducting pairing symmetry.  In this letter, we describe neutron scattering data that confirm for the first time the existence of a resonant spin excitation below T$_c$ in the iron arsenide materials at an energy that shows similar scaling to other heavy fermion and high-T$_c$ superconductors.

\begin{figure}[!hb]
\centering
\includegraphics[width=0.8\columnwidth]{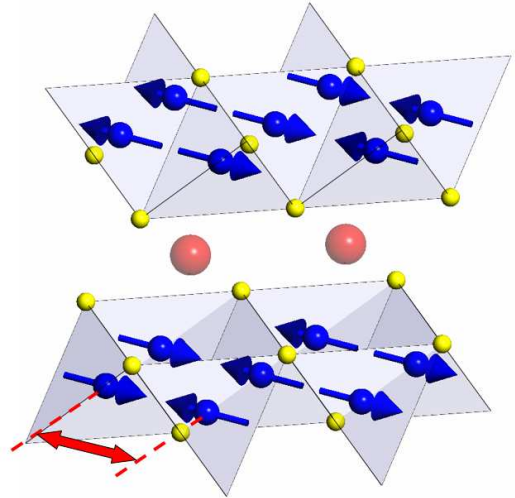}
\caption{The crystal structure of Ba$_{0.6}$K$_{0.4}$Fe$_2$As$_2$ (Fe: blue spheres, As: yellow spheres, Ba/K: red spheres).  The unit cell contains two layers of Fe$_2$As$_2$ tetrahedra separated by planes of barium atoms.  The blue arrows show the ordering of the iron spins observed in the undoped parent compound BaFe$_2$As$_2$ \cite{Huang:2008p8145}.  The atomic distance of 2.77\,\AA\  that characterizes both the antiferromagnetic modulation and the newly observed resonant excitation is shown by the red arrows.
\label{Fig1} }
\end{figure}

\begin{figure*}[!ht]
\centering
\includegraphics[width=1.95\columnwidth]{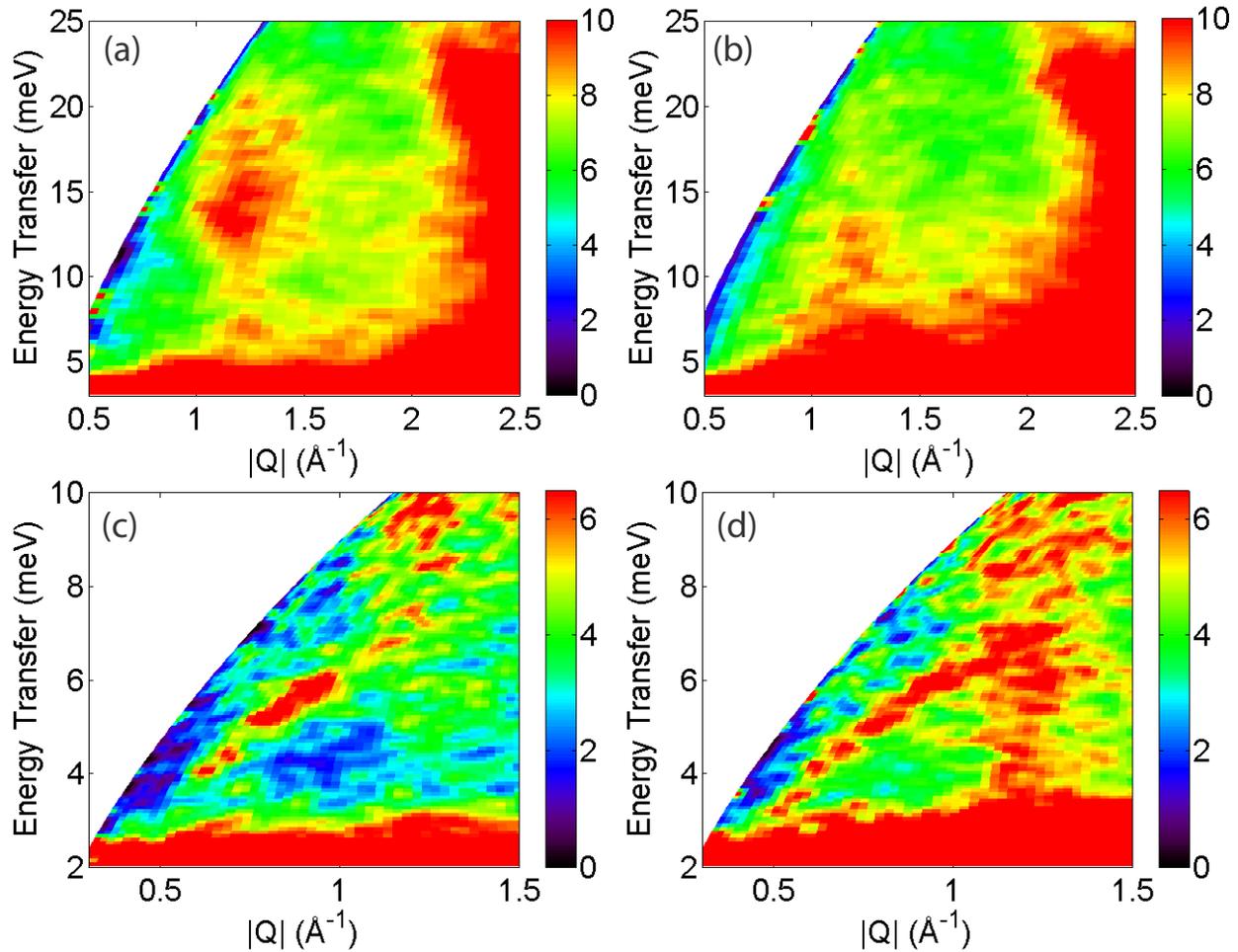}
\caption{Inelastic neutron scattering from Ba$_{0.6}$K$_{0.4}$Fe$_2$As$_2$ measured using incident neutron energies of 60\,meV (a,b) and 15\,meV (c,d), at temperatures of 7\,K (a,c) and 50\,K (b,d).  Above T$_c$ (b,d), the magnetic response consists of a column of excitations centred at 1.15\,\AA$^{-1}$ and a dispersive ferromagnetic excitation.  Below T$_c$, low-energy intensity in the column is transferred to the resonant excitation at 15\,meV.  The strong scattering at low energy transfers in each plot arises from the tail of strong elastic nuclear scattering, while the scattering increases strongly at higher Q due to inelastic phonon scattering. The colour scale is in units of mbarns/sr/meV/mol.  
\label{Fig2} }
\end{figure*}

Although the first iron arsenide superconductors were based on doped variants of  \textit{R}FeAsO, where \textit{R} is a rare earth element, there has been considerable interest in a new series of tetragonal compounds based on \textit{A}Fe$_2$As$_2$ (\textit{A} = Ba, Sr, Ca), in which superconductivity is induced either by doping the \textit{A}-site with potassium or sodium \cite{Rotter:2008p8535,Sasmal:2008p8092} or by applying pressure \cite{Park:2008p8883}.  These contain the same tetrahedrally-coordinated Fe$_2$As$_2$ planes as the LaFeAsO compounds (Fig. 1), separated by planes of the doped \textit{A}-site, which acts as a charge reservoir.  So far, the maximum T$_c$ is 38\,K \cite{Rotter:2008p8535} seen in Ba$_{0.6}$K$_{0.4}$Fe$_2$As$_2$, which is the compound we are investigating in this letter.  The antiferromagnetic structure of the undoped parent compound, BaFe$_2$As$_2$ is illustrated in Figure 1 \cite{Huang:2008p8145}.

Polycrystalline samples of Ba$_{0.6}$K$_{0.4}$Fe$_2$As$_2$ were prepared by solid state synthesis techniques described in the supplemental information. From XRD and SEM/EDS measurements, the estimated phase purity of the samples is $\sim$90\% and a sharp superconducting transition was observed by magnetic susceptibility at the previously reported temperature of 38\,K \cite{Rotter:2008p8535}.  The inelastic neutron scattering experiments were performed on the recently commissioned time-of-flight MERLIN spectrometer at the ISIS Pulsed Neutron and Muon Facility \cite{Bewley:2006p11347}, using incident energies of 15, 30, 60, and 100\,meV.  The data were placed on an absolute intensity scale by normalization to a vanadium standard.  

Figure 2 shows colour plots of the measured inelastic neutron scattering intensity as a function of momentum transfer, $Q$, and energy transfer, $\omega$, at two incident neutron energies, 15 and 60\,meV, below (T = 7\,K) and above (T = 50\,K) the superconducting transition temperature.  The range of data is limited by the kinematics of the scattering process at these low scattering angles so it is necessary to combine data from multiple incident energies in order to access data over a reasonable range of energy transfers at such low $Q$s.  

The most striking difference between the scattering above and below T$_c$ is seen at \textit{Q}$\sim$1.15\,\AA$^{-1}$ and an energy transfer of $\sim$14\,meV.  At 7\,K, there is clearly a peak that is well-defined in both $Q$ and $\omega$ which is not present at 50\,K (Fig. 2a and 2b).  Conversely, measurements at lower incident energy show that above T$_c$, there is a column of scattering intensity as a function of energy transfer, centred at $Q=1.15$\,\AA$^{-1}$ (Fig. 2d), which is no longer visible below T$_c$ (Fig. 2c).  The $Q$ characterizing these contributions to the magnetic response corresponds to the periodicity of the antiferromagnetic order within each plane of iron spins observed in the undoped parent compound, BaFe$_2$As$_2$ \cite{Huang:2008p8145} (see also Fig. 1), so this scattering is consistent with the persistence of antiferromagnetic fluctuations in the absence of long-range spin order.  Our measurements indicate that these antiferromagnetic fluctuations condense into a sharp resonant excitation, localized in both Q and $\omega$, as previously seen in the cuprate and heavy fermion superconductors.

\begin{figure}[!t]
\centering
\includegraphics[width=0.95\columnwidth]{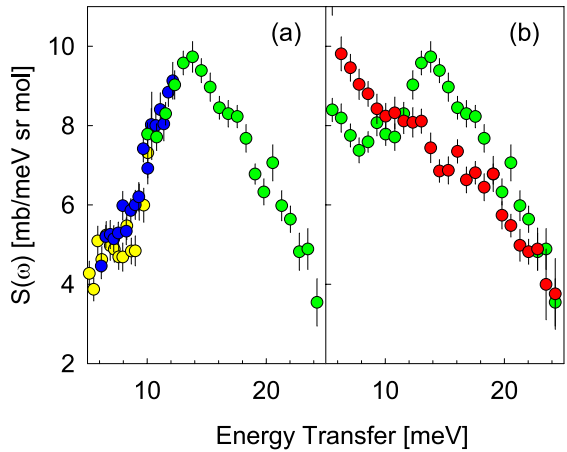}
\vspace{-0.2in}
\caption{Inelastic neutron scattering from Ba$_{0.6}$K$_{0.4}$Fe$_2$As$_2$ integrated over a Q-range of 1.0 to 1.3\,\AA$^{-1}$ (a) at 7\,K measured using incident neutron energies of 15\,meV (yellow circles), 30\,meV (blue circles) and 60\,meV (green circles), and (b) at 7\,K (green circles) and 50\,K (red circles) using an incident neutron energy of 60\,meV.  The error bars are derived from the square root of the raw detector counts.  The data show a clear resonant peak at 7\,K and the transfer of spectral weight from this peak to lower energies at 50\,K, \textit{i.e.}, above the superconducting transition.
\label{Fig3} }
\end{figure}

Before discussing this resonant excitation in more detail, we should mention that there is also clear evidence for ferromagnetic fluctuations shown by the additional band of excitations in Fig. 2c and 2d that disperses in Q and $\omega$.  The (Q,$\omega$)-dependence is consistent with the dispersion of ferromagnetic spin waves emerging from Q = 0, flattening at the ferromagnetic Brillouin zone boundary (which is equivalent to the antiferromagnetic zone centre) at $\omega\sim6$\,meV.  The existence of ferromagnetic fluctuations in the undoped parent compound has been predicted theoretically \cite{Mazin:2008p6770}, although it was argued that they would be suppressed by doping before the onset of superconductivity.  We have also observed this mode in BaFe$_2$As$_2$ but it is clear that they are still present in the superconducting phase although there is some redistribution of intensity on crossing T$_c$.  The strong intensity of this mode, coupled with the intensity changes at T$_c$, indicate that it is intrinsic to the bulk superconducting phase.   We will discuss this aspect of our data further in a future publication.

To elucidate the evolution of the magnetic response, we combine measurements below T$_c$ at three incident energies using, in each case, a low-energy cutoff that excludes the tail of strong elastic nuclear scattering.   The resulting data are shown in Fig. 3a where the resonant excitation is seen to peak sharply at $\omega_0\approx14$\,meV.  We cannot rule out that there could be a small phononic contribution to the scattering within this energy window, but the strong temperature dependence indicates that it is predominantly magnetic.  A comparison of the data above and below T$_c$ measured with an incident energy of 60\,meV shows that the spectral weight of the column of scattering evident in Fig. 2d has transferred into the resonant peak in the superconducting phase.

\begin{figure}[!b]
\centering
\includegraphics[width=0.75\columnwidth]{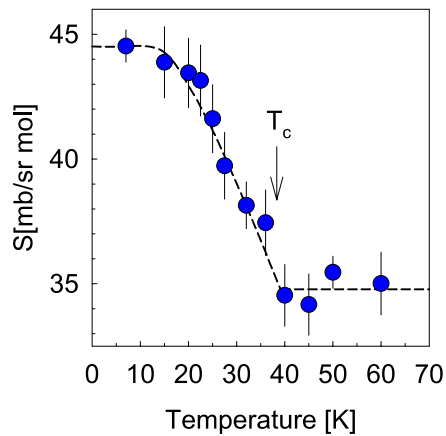}
\vspace{-0.2in}
\caption{Inelastic neutron scattering from Ba$_{0.6}$K$_{0.4}$Fe$_2$As$_2$ as a function of temperature integrated over a $Q$ range of 1.0 to 1.3\,\AA$^{-1}$ and an energy transfer range of 12.5 to 17.5\,meV. The integration range corresponds to the region of maximum intensity of the resonant excitation observed below T$_c$ (see Fig. 2).  The error bars are derived from the square root of the raw detector counts.  The dashed line is a guide to the eye below T$_c$ and shows the average value of the integrals above T$_c$.
\label{Fig4} }
\end{figure}

We performed a series of shorter measurements in order to determine the temperature dependence of this resonant excitation.  Fig. 4 shows data integrated over the (Q,$\omega$)-region of maximum intensity in the resonant excitation.  As also observed in the cuprates, the intensity of the resonance falls to zero at T$_c$ confirming the strong coupling of this excitation to the superconducting order parameter.   

The existence of similar resonant excitations in other strongly correlated superconductors, such as the high-T$_c$ cuprates and the heavy fermion superconductors, is commonly taken as evidence of an unconventional symmetry of the superconducting order parameter \cite{Chang:2007p11651}.  The dynamic magnetic susceptibility is predicted to be enhanced at certain values of \textbf{Q} in the superconducting phase by a coherence factor provided that the energy gap symmetry has the following form:

\[
\Delta_{\textbf{k}+\textbf{Q}} = -\Delta_{\textbf{k}}
\]

where $\textbf{k}$ and $\textbf{k}+\textbf{Q}$ are wavevectors on different parts of the Fermi surface.

In the cuprates and heavy fermion superconductors, this is realized by  \textit{d}$_{x^2-y^2}$-symmetry, which has nodes in the energy gap within a single Fermi surface.  In these cases, \textbf{Q} spans sections of the same Fermi surface that are gapped with opposite phase, such as $\textbf{Q}=(\pi,\pi)$ in the cuprates.  However, such a scenario seems to be ruled out by photoemission results on Ba$_{0.6}$K$_{0.4}$Fe$_2$As$_2$ that show no evidence of any anisotropy of the energy gap \cite{Ding:2008p8816}.  According to band structure calculation, the Fermi surfaces of the iron arsenide superconductors are predominantly derived from the iron \textit{d}-electrons, and comprise two small hole pockets centred at the centre of the Brillouin zone and two small electron pockets at the zone boundary \cite{Liu:2008p8202,Yang:2008p8157}.  ARPES sees isotropic gaps around each of the measured surfaces apparently ruling out a \textit{d}-wave gap symmetry \cite{Ding:2008p8816}.

A resolution of this apparent discrepancy has been provided by theoretical predictions that the gap symmetry is not \textit{d}-wave, but rather extended \textit{s$_\pm$}-wave \cite{Mazin:2008p6770}.  Mazin \textit{et al} have postulated that the gaps in each hole and electron pocket are isotropic, but that the gaps on the hole and electron pockets have opposite phase.  This means that magnetic fluctuations  are amplified by the coherence factor at values of \textbf{Q} that couple the hole and electron pockets, as has been confirmed by explicit calculations of the neutron scattering intensities \cite{Maier:2008p9552}.  This is precisely where we have observed the resonant excitation, so our measurements, combined with the ARPES data, provide strong experimental support for the validity of extended \textit{s$_{\pm}$}-wave gap models.

In conclusion, we have presented the first experimental evidence in the new class of iron arsenide superconductors for the existence of a resonant excitation in the dynamic magnetic susceptibility that disappears above the superconducting transition temperature.  The energy of this resonant excitation is at $\omega_0\sim14$\,meV, or 4.3\,T$_c$, just under the canonical value of 5\,T$_c$ seen in the cuprate superconductors \cite{Hufner:2008p9494}.  However, Stock \textit{et al} have argued that it is more appropriate to scale $\omega_0/2\Delta_0$, where $\Delta_0$ is the maximum value of the gap, and they estimate that this ratio ranges from 0.62 to 0.74 in a wide range of materials  \cite{Stock:2008p6655}. From ARPES data on Ba$_{0.6}$K$_{0.4}$Fe$_2$, $\Delta_0\sim12$\,meV \cite{Ding:2008p8816}, giving a ratio of  $\omega_0/2\Delta_0\sim0.58$.  It is remarkable that materials with such a divergent range of T$_c$s (over two orders of magnitude) could be unified by such a simple scaling relation.

\begin{acknowledgments}
We acknowledge helpful scientific discussions with Christopher Stock.  This work was supported by the Division of Materials Sciences and Engineering Division and the Scientific User Facilities Division of the Office of Basic Energy Sciences, U.S. Department of Energy Office of Science, under Contract Nos. DE-AC02-06CH11357 and DE-AC05-00OR22725. 

Correspondence and requests for materials should be addressed to R. O.

\end{acknowledgments}
%\bibliography{BaFe2As2_resonance}

\end{document}